\documentclass[runningheads]{llncs}
\usepackage[T1]{fontenc}
\usepackage{amsmath}
\usepackage{graphicx}
\usepackage{booktabs}
\usepackage{multirow}
\usepackage{amssymb}
\usepackage{hyperref}
\usepackage[symbol]{footmisc}
\begin{document}
\title{Subspace Implicit Neural Representations for Real-Time Cardiac Cine MR Imaging}
\titlerunning{Subspace INRs for Real-Time Cardiac Cine MR Imaging}
%

\author{Wenqi Huang\inst{1} \and
Veronika Spieker\inst{1,2} \and
Siying Xu\inst{3} \and
Gastao Cruz\inst{4} \and
Claudia Prieto\inst{5,6} \and
Julia Schnabel\inst{1,2,6} \and
Kerstin Hammernik\inst{1} \and
Thomas Küstner\inst{3} \and
Daniel Rueckert\inst{1,7}
}
\authorrunning{W. Huang et al.}

%
\institute{
Technical University of Munich, Germany \and 
Helmholtz Munich, Germany \and
University of Tübingen, Germany \and
University of Michigan, United States \and
Pontificia Universidad Católica de Chile, Chile \and
King’s College London, United Kingdom \and
Imperial College London, United Kingdom\\
\email{wenqi.huang@tum.de}
}
%
\maketitle              
\begin{abstract}




Conventional cardiac cine MRI methods rely on retrospective gating, which limits temporal resolution and the ability to capture continuous cardiac dynamics, particularly in patients with arrhythmias and beat-to-beat variations. To address these challenges, we propose a reconstruction framework based on subspace implicit neural representations for real-time cardiac cine MRI of continuously sampled radial data. This approach employs two multilayer perceptrons to learn spatial and temporal subspace bases, leveraging the low-rank properties of cardiac cine MRI. Initialized with low-resolution reconstructions, the networks are fine-tuned using spoke-specific loss functions to recover spatial details and temporal fidelity. Our method directly utilizes the continuously sampled radial \textit{k}-space spokes during training, thereby eliminating the need for binning and non-uniform FFT. This approach achieves superior spatial and temporal image quality compared to conventional binned methods at the acceleration rate of 10 and 20, demonstrating potential for high-resolution imaging of dynamic cardiac events and enhancing diagnostic capability.\footnote[1]{Code available upon acceptance: 
\url{https://github.com/wenqihuang/SubspaceINR-CMR}}

\end{abstract}
\section{Introduction}

Cardiac cine magnetic resonance imaging (MRI) plays an important role in accurate clinical evaluation of cardiovascular function and morphology, offering a non-invasive way to capture the beating heart. However, the presence of both cardiac and respiratory motion makes it challenging to acquire and reconstruct high-quality images, as achieving high spatial and temporal resolution simultaneously remains difficult~\cite{spieker2023deep}. Therefore, fast data acquisition with high undersampling rates in the MR signal space, also called \textit{k}-space, has attracted great interest in the field of reconstruction.  

Reconstruction of images from undersampled \textit{k}-space data has been significantly advanced by the introduction of parallel imaging (PI) and compressed sensing (CS) techniques in MRI. These approaches have driven substantial improvements in acquisition efficiency, leveraging both hardware and algorithmic innovations. PI utilizes multiple receiver coils in MRI scanners, exploiting the redundancy among coils to enable higher undersampling rates during data acquisition. In contrast, CS incorporates sparse priors as regularization terms
to constrain the solution space and mitigate the risk of local minima in the reconstruction process. 
Building upon these foundations, the development of low-rank and subspace methods has opened up new avenues for dynamic cardiac MRI reconstruction~\cite{lingala2011accelerated,christodoulou2014improved,otazo2015low,feng2020grasp}. By explicitly leveraging the low-rank structure across temporal phases, these approaches model the cardiac motion as lying close to a lower-dimensional manifold, which can be spanned by just a few temporal basis functions. This effective dimensionality reduction not only simplifies the reconstruction problem but also enables the use of fewer measurements, ultimately improving scan efficiency. As a result, both spatial and temporal information are faithfully recovered, facilitating more accurate representations of the underlying cardiac motion.
More recently, deep learning-based approaches have demonstrated remarkable potential in leveraging acquisition physics and raw \textit{k}-space data for improved reconstruction~\cite{schlemper2017deep,hammernik2018learning,qin2018convolutional,huang2021lsnet,ke2021learned,pan2024reconstruction,pan2024unrolled}. Despite their success, most of these methods are supervised and require large, fully-sampled training datasets, and consequently prone to hallucinations when there are domain shifts. 
Several approaches that do not require fully sampled data have been proposed~\cite{akccakaya2019scan,yaman2020self,hu2021self,desai2023noise2recon,korkmaz2023self,cui2022self}, and while recent work has begun to address radial sampling~\cite{blumenthal2024self,mancu2023self}, the majority of these methods have been developed primarily for Cartesian sampling trajectories.

Non-Cartesian sampling techniques, such as radial sampling, have become increasingly popular. Unlike Cartesian sampling, non-Cartesian approaches vary both phase encoding directions simultaneously, which helps to distribute motion- and undersampling-related noise throughout the image, leading to more incoherent and less prominent artifacts~\cite{wright2014non}. This property makes non-Cartesian sampling particularly well-suited for cardiac cine MRI, especially in patients with irregular heart rhythms. However, non-Cartesian sampling introduces new challenges. Reconstruction from non-Cartesian data requires non-uniform fast Fourier transforms (NUFFT), which can be computationally expensive and affected by density compensation and interpolation errors~\cite{pipe1999sampling,wright2014non}. 
Due to the moving nature of the heart and the limitations of MR physics, only a limited number of data points can be rapidly sampled for each motion state in cardiac cine MRI. Existing methods typically rely on binning \textit{k}-space samples along the temporal dimension into a fixed number of motion states to achieve high image quality for each cardiac frame~\cite{ahmad2015variable,jung2010radial,rajiah2023cardiac,wang2021fast}. Binning of real-time data simplifies data representation and optimization, but results in a loss of temporal information.

In recent developments, Implicit Neural Representations (INRs) have shown impressive potential to represent continuous signals through coordinate-value mappings using multi-layer perceptrons (MLPs)~\cite{chibane2020implicit,sitzmann2020implicit,mildenhall2021nerf,muller2022instant}. The flexibility of INRs to sample at arbitrary locations makes them ideal to reconstruct images from non-Cartesian sampling trajectories without the need for NUFFT, and the ability of continuous representation is ideal for learning the continuous spatiotemporal information in cardiac cine MRI. Previous studies show the effectiveness of INRs for both Cartesian~\cite{feng2022spatiotemporal,shen2022nerp,kunz2024implicit} and non-Cartesian sampling~\cite{catalan2023unsupervised}, but cardiac cine MRI methods still either rely on binned data~\cite{catalan2023unsupervised} or attempt to directly learn the entire spatiotemporal space~\cite{huang2023neural,spieker2023iconik,haft2024neural,spieker2024self}, which is particularly challenging under sparse sampling schemes.

Despite significant progress in cardiac cine MRI reconstruction, achieving high-quality images for real-time cardiac cine MRI under sparse sampling conditions remains a challenging task. Traditional binning-based approaches sacrifice temporal resolution by grouping data into discrete motion states, while non-Cartesian sampling often relies on computationally expensive and error-prone NUFFT operations. Moreover, directly modeling the full spatiotemporal domain of the beating heart can overwhelm reconstruction algorithms, particularly when acquired data are limited. To overcome these hurdles, we propose a novel approach that exploits the inherent low-rank structure of cardiac cine MRI through the integration of subspace learning and INRs. By avoiding binning and NUFFT altogether, our approach enables efficient, high-quality reconstruction that preserves crucial cardiac dynamics. Our main contributions are as follows:

\begin{enumerate}
    \item We introduce a reconstruction pipeline that tailored for radially sampled real-time cardiac cine MRI that removes the need for temporal binning and avoids the limitations of NUFFT.
    \item We propose a subspace learning strategy that directly exploits the low-rank properties of cardiac cine data. By representing the data as a combination of spatial and temporal bases, we effectively reduce the dimensionality of the representation space. This reduction enables our approach to employ INRs to accurately learn these representations, thereby capturing intricate details of cardiac motion.
    \item Our framework achieves superior reconstruction of both spatial details and temporal motion compared to conventional binning-based methods, delivering improved image quality and more faithful depictions of cardiac function.
\end{enumerate}

\section{Methods}

This section presents our proposed reconstruction framework for real-time cardiac cine MRI, integrating INRs and subspace learning. By leveraging the low-rank properties of cardiac cine MRI and bypassing the need for binning and NUFFT, our method achieves superior spatiotemporal fidelity while directly utilizing continuous \textit{k}-space sampling. The following subsections detail the problem formulation, the INR-based reconstruction strategy, and the integration of low-rank properties for efficient representation.

\begin{figure}[h]
    \centering
    \includegraphics[width=\linewidth]{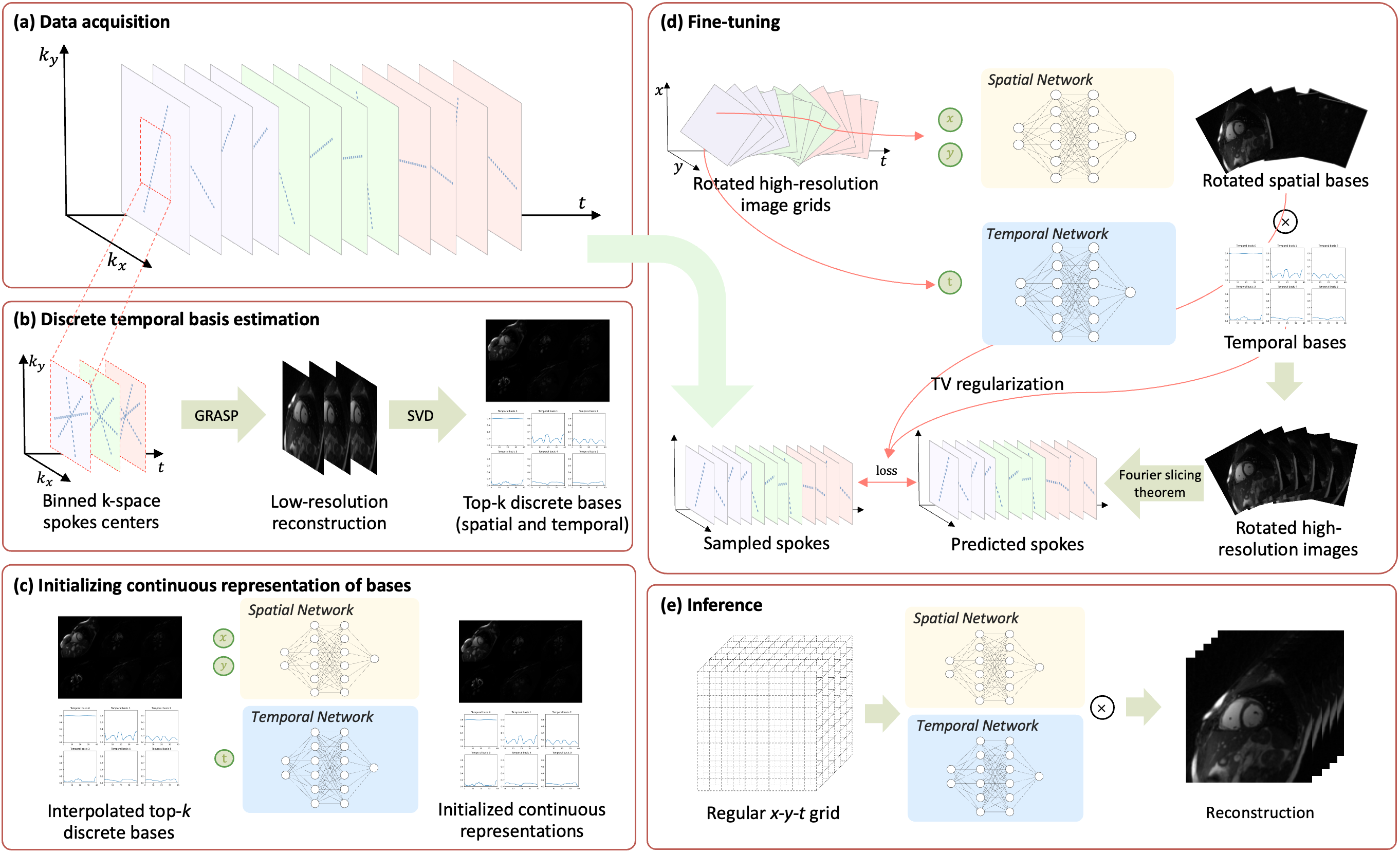}
    \caption{
    Method overview: a) Continuous \textit{k}-space sampling with tiny golden angle radial trajectory; b) Binning spoke centers to reconstruct a low-resolution image with GRASP, decomposing into spatial and temporal bases via SVD, retaining top-k components; c) Interpolating and fitting low-resolution bases to spatial and temporal networks; d) Inputting rotated spatial and accurate temporal coordinates into networks to obtain spatial and temporal bases, whose product forms rotated images, networks optimized with Eq. \ref{eq:fine_tune}; e) Final reconstruction by inputting a regular $x-y-t$ grid.
    }
    \label{fig:diagram}
\end{figure}

\subsection{Cardiac Cine MRI with Continuous Radial Sampling}
The raw signal of cardiac cine MRI with continuous radial sampling trajectories is acquired for each spoke in the \textit{k}-space (frequency domain). The goal of MRI reconstruction is to recover the optimal image $\mathbf{x}^*$ by solving:
\begin{equation}
    \mathbf{x}^* = \arg\min_\mathbf{x} DC(\mathbf{x}, \mathbf{y}) + R(\mathbf{x}).
\end{equation}
Here $DC(\mathbf{x}, \mathbf{y})$ enforces data consistency between the reconstructed image $\mathbf{x}$ and the \textit{k}-space measurements $\mathbf{y}$, and $R(\mathbf{x})$ is a regularization term. Each spoke $\mathbf{x}_i, i=1,2,\cdots,N$ ($N$ for total number of spokes), sampled at time intervals defined by the repetition time (TR), represents \textit{k}-space information from a single moment in time. However, it is impossible to reconstruct an image frame from a single spoke due to the extreme sparsity of information. Conventionally, neighboring \textit{k}-space spokes (in time) are grouped into segments using a binning operator $B$, and the data consistency term is defined as:
\begin{equation}
\label{eq:dc_conventional}
    DC(\mathbf{x}, \mathbf{y}) = ||A \mathbf{x} - B \mathbf{y}||^2_2,
\end{equation}
where $A = MFS$, representing the undersampling mask $M$, NUFFT operator $F$, and coil sensitivity maps $S$. To encounter the lack of measurements per time point and the resulting undersampling in the spatial domain, the binning operator $B$ aggregates neighboring spokes from various time points (see various colors in \ref{fig:diagram}a). For larger bin sizes, more cardiac motion is averaged within one frame, which in turn can lead to an increased motion blurring in the reconstructed image. Therefore, there will be an unavoidable trade-off between temporal and spatial resolution when using binning. 

\subsection{Implicit Neural Representations for MRI Reconstruction}
To fully utilize the temporal information encoded in the acquired \textit{k}-space spokes, an ideal approach would reconstruct $N$ image frames, each corresponding to a single spoke, eliminating the temporal information loss from the binning operator $B$. This modifies the data consistency term to:
\begin{equation}
    DC(\mathbf{x}, \mathbf{y}) = ||A \mathbf{x} - \mathbf{y}||^2_2.
\end{equation}
However, the direct application of NUFFT to each spoke results in suboptimal reconstructions. To avoid that, we represent the image $\mathbf{x}$ as a continuous function using an INR network, $\mathbf{x} = G_{\theta}(\mathbf{c})$, where $\mathbf{c}=\{c_i=[x_i, y_i, t_i]^{\top}, i=1,2,\cdots,M\}$ denotes coordinates in the spatiotemporal space, and $G_{\theta}(\cdot): \mathbb{R}^3 \rightarrow \mathbb{C}$ is the INR network mapping from the spatiotemporal coordinates to the complex image voxel value. The optimal reconstruction is then obtained by optimizing the network parameters $\theta$:
\begin{equation}
    \begin{split}
        \theta^* &= \arg\min_{\theta} DC(G_{\theta}(\mathbf{c}), \mathbf{y}) + R(G_{\theta}(\mathbf{c})), \\
        \mathbf{x}^* &= G_{\theta^*}(\mathbf{c}_{\text{grid}}),
    \end{split}
    \label{eq:inr_problem_formulation}
\end{equation}
where $\theta^*$ represents the optimal parameters of the INR network, $\mathbf{c}_{\text{grid}}$ represents the coordinates of a target MRI image on a Cartesian grid.

The sparsity of \textit{k}-space, as well as the high-dimensional $x$-$y$-$t$ image space, poses a challenge for learning a well-represented INR. To address this, we leverage the inherent temporal redundancies of cardiac cine MRI by integrating a compact subspace into the INR training. Inspired by the low-rank representations of cardiac cine MRI, where the dynamic image can be approximated by the product of two low-rank matrices (e.g., top-$k$ components) — spatial bases and temporal bases - we decompose the dynamic image into these two components. The image is decomposed into spatial and temporal bases, modeled by two networks: $G_{\theta_s}(\mathbf{s})$ for spatial representation and $G_{\theta_t}(\mathbf{t})$ for temporal representation. The optimization problem is reformulated as:
\begin{equation}
\begin{split}
    \theta_s^*, \theta_t^* &= \arg\min_{\theta_s, \theta_t} DC(G_{\theta_s}(\mathbf{s}) \cdot G_{\theta_t}(\mathbf{t})) + R_s(G_{\theta_s}(\mathbf{s})) + R_t(G_{\theta_t}(\mathbf{t})), \\
    \mathbf{x}^* &= G_{\theta_s^*}(\mathbf{s}) \cdot G_{\theta_t^*}(\mathbf{t}).
\end{split}
\label{eq:fine_tune}
\end{equation}
Here $\theta_s^*$ and $\theta_t^*$ represent the optimal parameters for the spatial and temporal model, respectively, while $R_s$ and $R_t$ denote their networks' corresponding regularization terms; $\mathbf{s}$ are the spatial coordinates and $\mathbf{t}$ represents the temporal coordinates. The spatial model $G_{\theta_s}(\cdot):\mathbb{R}^2\rightarrow\mathbb{C}^k$ maps the spatial coordinates to a $k$-dimensional output for spatial bases, and the temporal model $G_{\theta_t}(\cdot): \mathbb{R}\rightarrow\mathbb{C}^k$ maps the temporal coordinates to a $k$-dimension output. This formulation reduces the parameter space while preserving essential spatiotemporal information, enabling efficient and accurate reconstruction.

Leveraging the INR's ability to sample arbitrary coordinates, we use the Fourier slice theorem~\cite{bracewell1956strip,Nishimura2010} to establish a direct relationship between the image and its \textit{k}-space spokes. The theorem states that the 1D Fourier transform of a line integral (projection) $p_v(t)$ of a 2D image $f(x,y)$ along a certain direction $v$ is equivalent to a corresponding spoke in \textit{k}-space:
\begin{equation}
    \mathcal{F}_1\{p_v(t)\}(\omega) = \mathcal{F}_2\{f(x, y)\}(\omega_x, \omega_y).
\end{equation}
Here $(\omega_x, \omega_y)$ lies along the spoke's direction (vertical to projection direction $v$). 
This means that, in our case, each spoke corresponds to the 1D Fourier transform of the image's projection along the spoke's vertical direction,
\begin{equation}
    \mathcal{F}_1 P_{v}(S\mathbf{x}) = \mathbf{y},
\end{equation}
where $P_v$ is the projection operator for the image $\mathbf{x}$ along the vertical direction of each spoke, and $S$ represents the coil sensitivities. This theorem establishes a direct relationship between radial spokes in $k$-space and the image space without requiring NUFFT. Furthermore, when combined with INR, it eliminates the need for image interpolation when applying the projection $P_v$ to handle rotated grid coordinates in the problem formulation (Eq.~\ref{eq:inr_problem_formulation}), as INRs can directly sample points on a rotated image grid. The data consistency term is reformulated as:
\begin{equation}
    DC(G_{\theta_s}(\mathbf{s}) \cdot G_{\theta_t}(\mathbf{t}), \mathbf{y}) = w (\mathcal{F}_1P_v(S\cdot G_{\theta_s}(\mathbf{s}) \cdot G_{\theta_t}(\mathbf{t})) - \mathbf{y})
\end{equation}
Here $w$ is a ramp weight, which is the distance to the spoke center of each data points, compensating the value range of low and high frequency area in $k$-space spokes. By sampling on a rotated image grid according to the rotation angle of each spoke, the projection simplifies to a summation along the vertical direction to each spoke. This allows for direct utilization of continuous \textit{k}-space spokes without binning, as inspired by \cite{catalan2023unsupervised}.


\subsection{Network Initialization and Optimization}
Proper initialization is critical for fast convergence. Inspired by \cite{feng2020grasp,qiu2024self}, we first reconstruct a low-resolution image using GRASP~\cite{feng2014golden}, aggregating the central spokes of low spatial and temporal resolution. This low-resolution image is decomposed via singular value decomposition (SVD), retaining the top-$k$ components (Figure~\ref{fig:diagram}b), which are interpolated to full resolution using linear interpolation. The spatial and temporal networks are initialized by minimizing the mean squared error (MSE) between their outputs and these interpolated bases (Figure~\ref{fig:diagram}c). Subsequently, the network parameters are optimized by minimizing Eq.~\ref{eq:fine_tune}. Once trained, the reconstructed image is obtained by inputting the $x$-$y$-$t$ grid $\mathbf{c}_{\text{grid}}$ into the networks. Figure~\ref{fig:diagram}d and e illustrate the fine-tuning and inference steps of the proposed method.

\section{Experimental Setup}
\subsubsection{Dataset}
Experiments were conducted on 17 healthy subjects using a 1.5T scanner (Ingenia, Philips, Best, The Netherlands) with 28 MR receiver coils. To reduce training time, we used data from 6 selected coils sensitive to the heart region. 
All experiments were IRB-approved with written informed consent. 
The following sequence parameters were used: 7-8 short axis slices; FOV = $256\times256$ mm$^2$ (considering inherent $1.6\times$ frequency encoding oversampling of radial trajectories); $8$ mm slice thickness; resolution = $2\times2$ mm$^2$; TE/TR = $1.16/2.3$ ms; b-SSFP readout; radial tiny golden angle of ${\sim}23.6^\circ$; flip angle $60^\circ$; 8960 acquired radial spokes (800 spokes were used, covering 2-3 cardiac cycles) without ECG gating; nominal scan time ${\sim}20$s (${\sim}1.8$s for 800 spokes); breath-hold acquisition. No fully-sampled ground truth images are available in this real-time acquisition setting.

\subsubsection{Model Details} The proposed INR framework consists of a spatial network and a temporal network. Both networks are based on multi-layer perceptrons (MLPs) with hash-grid encoding~\cite{muller2022instant}. The hash-grid encoding uses a hashmap of size 20 with 16 levels, 2 features per level, a base resolution of 16, and a per-level scale of 1.26. The MLPs have 2 hidden layers, each containing 64 nodes. The number of main components $k$ is set to 6 for both spatial and temporal bases. The networks generate an output dimension of 12, which represents the concatenation of the real and imaginary parts of the basis components.

\subsubsection{Initialization Settings} The low-resolution image (1/2.56 of the original resolution) was reconstructed using GRASP~\cite{feng2014golden,uecker2016bart} with 100 iterations on binned spoke centers. Binning was performed with 20 spokes per phase and a total variation regularization weight of 0.025. By cropping the center of the radial \textit{k}-space for the low-resolution reconstruction, sparsity towards the outside \textit{k}-space can be reduced. Thus, the effective acceleration factor is reduced, resulting in low-resolution reconstructions with minimized undersampling artefacts~\cite{feng2022spatiotemporal}. SVD was applied to the image, decomposing it into spatial bases and temporal bases. Then we retained the top 6 components for further processing. The temporal and spatial bases were interpolated to resolutions of 2.56 times and 20 times their original sizes, respectively, to match the full spatial resolution and maximum temporal resolution (TR). The spatial and temporal networks were trained on these interpolated bases for 1000 steps with a learning rate of $0.01$, using MSE loss.

\subsubsection{Fine-Tuning Settings} Training on non-binned spokes was conducted using the Adam optimizer for 150 iterations on all sampled spokes at a reduced learning rate of $3\times10^{-5}$ for this phase. During the first 10 iterations, the temporal model was frozen to accelerate the optimization of the spatial model. The entire initiaslization and fine-tuning process for each slice took approximately 3 minutes on an NVIDIA RTX A6000 GPU with 48GB of memory. 

\subsubsection{Evaluation Settings} We compared our proposed method quantitatively and qualitatively to NUFFT and GRASP. Since NUFFT and GRASP require data binning, we configured them to use 20 and 40 spokes per bin (R=20 and R=10) to balance spatial and temporal resolution. For the proposed method, we sampled on the same image grid as the 20 spokes/bin configuration for a fair comparison. GRASP implementation was performed using the BART toolbox~\cite{uecker2016bart}. As ground truth images were unavailable, image quality was assessed quantitatively using edge sharpness (ES) and estimated signal-to-noise ratio (SNR) in the end-systolic and end-diastolic cardiac phases. ES is defined as the inverse distance between the positions corresponding to 20\% and 80\% of the maximum intensity along the line profile~\cite{shea2001coronary,kustner2020isotropic}. ES was measured by extracting 12 line profiles between the left ventricle blood pool and the endocardium (6 lines for each phase). A higher ES value indicates sharper edges. SNR was estimated by extracting patches from the background region (noise region) and the left ventricle blood pool (signal region). SNR was calculated as $10 \cdot \log_{10}(P_s / P_n)$, where $P_s$ and $P_n$ are the mean values of the signal and noise regions, respectively. Only the 2nd to 6th short-axis slices of each subject were considered in the ES and SNR calculations, as the other slices typically lacked sufficient information about the ventricular blood pool.

\section{Results} Table~\ref{tab:experiment_results} summarizes the mean and standard deviation of the quantitative results for systole and diastole phases. For SNR, the proposed method outperformed the comparison methods by a considerable margin in both systolic and diastolic phases, reflecting its ability to maintain a higher SNR during dynamic cardiac motion. For ES, the proposed method showed a notable advantage in the systolic phase and comparable performance in the diastolic phase, indicating its capability to preserve edge sharpness under different motion conditions.

\begin{table}[ht]
    \centering
    \caption{SNR and Edge Sharpness (ES) of comparison methods}
    \begin{tabular}{@{}lcccc@{}}
        \toprule
        \multirow{2}{*}{\textbf{Method}} & \multicolumn{2}{c}{\textbf{SNR (mean$\pm$std, dB)}} & \multicolumn{2}{c}{\textbf{ES (mean$\pm$std, $1/mm$)}} \\ 
        \cmidrule(lr){2-3} \cmidrule(lr){4-5}
        & \textbf{Systolic} & \textbf{Diastolic} & \textbf{Systolic} & \textbf{Diastolic} \\
        \midrule  
        NUFFT        & $6.85 \pm 2.45$  & $7.62 \pm 2.27$  & $0.177 \pm 0.024$ & $0.243 \pm 0.0287$ \\
        GRASP        & $13.57 \pm 2.79$ & $13.92 \pm 2.79$ & $0.201 \pm 0.030$ & $\mathbf{0.316 \pm 0.043}$ \\
        \textbf{Proposed} & $\mathbf{20.21 \pm 6.88}$ & $\mathbf{20.89 \pm 6.96}$ & $\mathbf{0.227 \pm 0.034}$ & $0.316 \pm 0.052$ \\ \bottomrule
    \end{tabular}
    \label{tab:experiment_results}
\end{table}

\begin{figure}[!h]
    \centering
    \includegraphics[width=\linewidth]{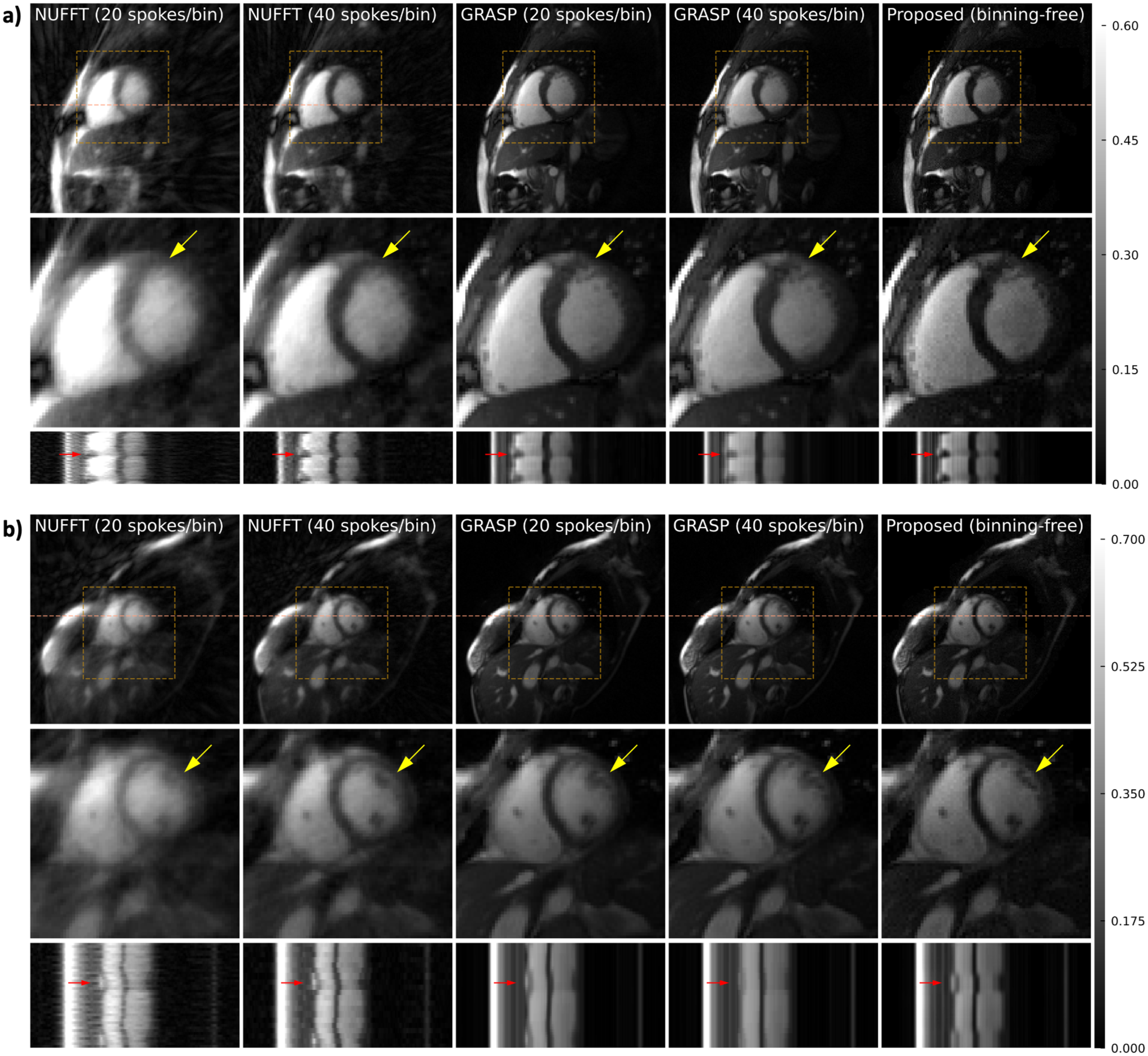}
    \caption{Comparison of reconstruction results. a) and b) show two subjects. The first row of each subfigure shows $x-y$ images at a selected cardiac phase. The second row zooms into a region of interest (ROI). The third row presents $x-t$ profiles along a chosen $y$-coordinate. Our method preserves details in the ROI (yellow arrow) and minimizes temporal blurring seen in binned methods (red arrow). The color bars indicate intensity ranges.}
    \label{fig:comparison}
\end{figure}

\begin{figure}
    \centering
    \includegraphics[width=0.8\linewidth]{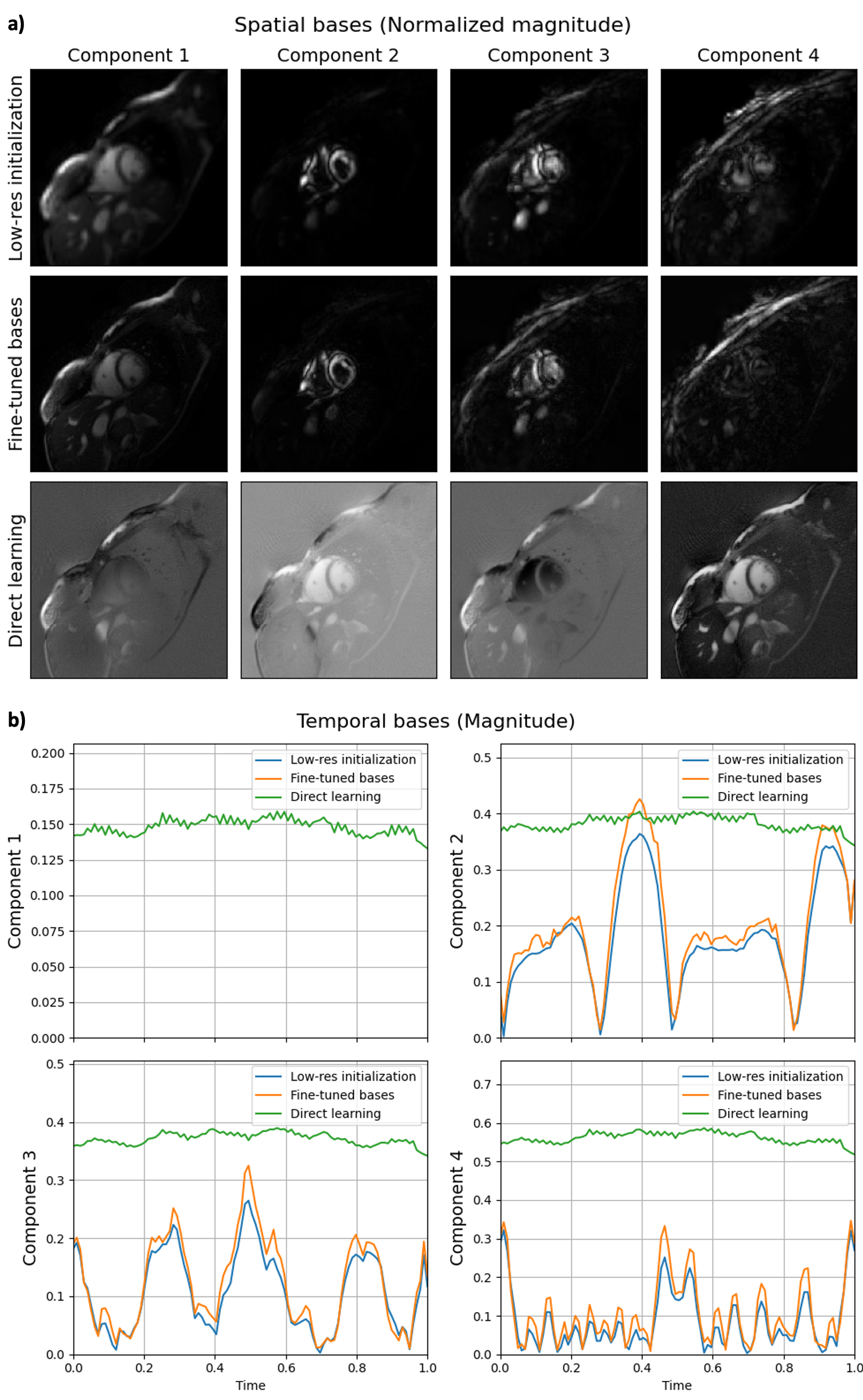}
    \caption{The first four components of the spatial and temporal bases for an example case. a) The spatial bases are shown across four columns, where the first row displays the initialized spatial bases derived from low-resolution reconstruction, the second row shows the fine-tuned representations obtained by training based on the low-resolution initialization, and the third row presents the bases directly learned without initialization. b) The temporal bases correspond to the spatial components in a), with orange lines representing the low-resolution initialization, green lines showing the fine-tuned bases, and blue lines depicting the directly learned components without initialization. The results demonstrate that with low-resolution initialization, the models successfully capture finer spatial structures and temporal details after training on non-binned \textit{k}-space spokes. In contrast, without initialization, the models fail to converge effectively.}
    \label{fig:bases_visualization}
\end{figure}

Figure~\ref{fig:comparison}a and b present qualitative comparisons between NUFFT, GRASP, and the proposed method for two representative cases. Higher temporal resolutions (fewer spokes per frame) are compared to higher spatial resolutions (more spokes per frame). Qualitative results demonstrate that the proposed method preserves finer spatial details and richer temporal fidelity compared to binned approaches, with negligible motion blur. In the region of interest, our method retains intricate details that are blurred in the reconstructions of binned methods (NUFFT and GRASP) marked by yellow arrows. This difference can be attributed to the averaging of neighboring time frames in binned methods, which introduces spatial and temporal blurring. The $x$-$t$ profiles further illustrate that binning leads to noticeable temporal blurring, while the proposed reconstruction framework effectively preserves temporal continuity, offering clear visualization of dynamic cardiac motion.


Figures~\ref{fig:bases_visualization}a and b illustrate the first four components of temporal and spatial bases learned by the networks under three different settings: low-resolution initialization only, fine-tuned results based on low-resolution initialization, and directly learned bases without low-resolution initialization. By comparing the results of directly learned and fine-tuned bases, we observe that without initialization, the INRs struggle to learn the spatiotemporal information, particularly failing to capture the temporal details. This highlights that initialization is crucial for achieving fast convergence and effective representation learning in our proposed method. Comparing the initialized and fine-tuned bases, we notice that the spatial bases initially appear blurry and lack fine details, as they are generated from low-resolution reconstructions during the initialization phase. After fine-tuning on the non-binned \textit{k}-space spokes, the spatial bases exhibit significantly enhanced clarity, and the temporal bases recover fine details that were previously lost. These results validate our hypothesis that the INR framework, combined with fine-tuning, effectively recovers high-frequency spatial and temporal information. This comparison further demonstrates the effectiveness of leveraging the low-rank property of cardiac cine MRI within the INR framework.

In summary, both qualitative and quantitative evaluations indicate that the proposed method achieves superior reconstruction quality compared to conventional NUFFT and GRASP methods. By preserving both spatial and temporal details, the proposed framework demonstrates significant advantages for the reconstruction of real-time cardiac cine MRI. The results suggest that this approach is particularly well-suited for capturing dynamic cardiac events, providing high-quality reconstructions without the need for binning or NUFFT.

\section{Discussion and Conclusion}


In this work, we presented a novel subspace INR framework for the reconstruction of real-time cardiac cine MRI. Conventional approaches often rely on binning \textit{k}-space data to handle motion. Similarily, they often rely on non-uniform FFT (NUFFT) to handle non-Cartesian sampling trajectories. While effective in certain scenarios, these methods face inherent limitations, such as losing temporal resolution due to binning and reconstruction artifacts arising from NUFFT density compensation and interpolation errors. Our framework addresses these challenges by leveraging subspace learning with INRs,  eliminating the need for both binning and NUFFT. Our proposed method exploits low-rank properties of cardiac cine MRI, enabling accurate and data-efficient reconstruction from sparse \textit{k}-space samples without compromising spatial or temporal resolution.

The foundation of our method lies in decomposing the spatiotemporal image into separate spatial and temporal bases represented by two distinct INR networks. This decomposition reduces the complexity of the reconstruction problem by projecting the high-dimensional spatiotemporal space into a lower-dimensional subspace, significantly improving data efficiency. The networks are initialized using low-resolution reconstructions obtained via GRASP, ensuring stable and reliable starting points. Fine-tuning networks with spoke-specific losses recovers fine details lost at initialization, resulting in superior image quality. Our experiments show that the proposed method outperforms traditional binning-based approaches, better preserving temporal dynamics and finer spatial details. Additionally, the flexibility of INR-based representations allows the reconstruction of continuous cardiac motion, effectively minimizing motion artifacts that are common in conventional methods.

One of the key contributions of this work is a paradigm shift in the pre-processing of \textit{k}-space data. Instead of treating groups of \textit{k}-space measurements, such as binned spokes, as the minimal unit, we explored treating individual spokes as the minimal unit. This approach unlocks the potential to achieve a temporal resolution equal to the repetition time (TR). While similar concepts have been explored for implicit neural representations in \textit{k}-space~\cite{huang2023neural}, our work extends the representation to image space while retaining the spoke-wise minimal unit through the use of the Fourier slicing theorem, showcasing its utility for real-time cardiac cine MRI. Furthermore, this concept is not limited to cardiac imaging, but can be generalized to multi-contrast imaging and other quantitative MRI applications, highlighting its broad potential impact.

Despite its very promising results, this study still has some limitations. First, while training and inference times have been reduced to approximately 15 minutes, this remains a bottleneck for clinical routine use. Second, the method requires training and 
additional optimization for each new scan, which prevents direct inference on unseen data. Although this scan-specific approach minimizes the risk of inpainting artifacts from training data, it introduces potential issues with overfitting. The current stopping criteria for the number of iterations are empirically determined and may not always be optimal. Lastly, our evaluation focused on imaging quality metrics; further studies are needed to assess the method's performance on more clinical parameters. These limitations underscore the need for continued refinement and validation.

Future work will focus on optimizing the spatial and temporal network architectures to improve computational efficiency and scalability. This is particularly important for extending the approach to higher-dimensional problems, such as 4D MRI, where the trade-off between data acquisition time and resolution is even more pronounced. Expanding the framework to other applications and imaging modalities, such as functional MRI or positron emission tomography (PET), could further demonstrate its versatility and applicability across the broader field of medical imaging. Finally, rigorous clinical validation, including evaluations on pathological cases and diverse multi-institution datasets, will be critical to ensure its practical utility in real-world clinical settings.

In conclusion, the proposed subspace INR framework represents a significant step forward in real-time cardiac cine MRI reconstruction. By overcoming the limitations of binning and NUFFT, our method preserves both spatial and temporal fidelity, enabling high-quality imaging of dynamic cardiac events. With further development and clinical validation, this approach has the potential to transform real-time cardiac cine MRI and expand its applications across a range of imaging scenarios.

\begin{credits}
\subsubsection{\ackname} This work was supported in part by the European Research Council (Deep4MI project, Grant Agreement Nr.884622). We would also like to thank Dr. Jing Cheng, Dr. Yuanyuan Liu and Dr. Zhilang Qiu for the insightful discussion during ISMRM 2024.

\end{credits}
%
%
%
\bibliographystyle{splncs04}
\bibliography{references}

\begin{thebibliography}{10}
\providecommand{\url}[1]{\texttt{#1}}
\providecommand{\urlprefix}{URL }
\providecommand{\doi}[1]{https://doi.org/#1}

\bibitem{ahmad2015variable}
Ahmad, R., Xue, H., Giri, S., Ding, Y., Craft, J., Simonetti, O.P.: Variable density incoherent spatiotemporal acquisition (vista) for highly accelerated cardiac mri. Magnetic resonance in medicine  \textbf{74}(5),  1266--1278 (2015)

\bibitem{akccakaya2019scan}
Ak{\c{c}}akaya, M., Moeller, S., Weing{\"a}rtner, S., U{\u{g}}urbil, K.: {Scan-specific robust artificial-neural-networks for k-space interpolation (RAKI) reconstruction: Database-free deep learning for fast imaging}. Magnetic resonance in medicine  \textbf{81}(1),  439--453 (2019)

\bibitem{blumenthal2024self}
Blumenthal, M., Fantinato, C., Unterberg-Buchwald, C., Haltmeier, M., Wang, X., Uecker, M.: Self-supervised learning for improved calibrationless radial mri with nlinv-net. Magnetic Resonance in Medicine  \textbf{92}(6),  2447--2463 (2024)

\bibitem{bracewell1956strip}
Bracewell, R.N.: Strip integration in radio astronomy. Australian Journal of Physics  \textbf{9}(2),  198--217 (1956)

\bibitem{catalan2023unsupervised}
Catal{\'a}n, T., Courdurier, M., Osses, A., Botnar, R., Costabal, F.S., Prieto, C.: Unsupervised reconstruction of accelerated cardiac cine mri using neural fields. arXiv preprint arXiv:2307.14363  (2023)

\bibitem{chibane2020implicit}
Chibane, J., Alldieck, T., Pons-Moll, G.: {Implicit functions in feature space for 3D shape reconstruction and completion}. In: Proceedings of the IEEE/CVF Conference on Computer Vision and Pattern Recognition. pp. 6970--6981 (2020)

\bibitem{christodoulou2014improved}
Christodoulou, A.G., Hitchens, T.K., Wu, Y.L., Ho, C., Liang, Z.P.: Improved subspace estimation for low-rank model-based accelerated cardiac imaging. IEEE Transactions on Biomedical Engineering  \textbf{61}(9),  2451--2457 (2014)

\bibitem{cui2022self}
Cui, Z.X., Cao, C., Liu, S., Zhu, Q., Cheng, J., Wang, H., Zhu, Y., Liang, D.: Self-score: Self-supervised learning on score-based models for mri reconstruction. arXiv preprint arXiv:2209.00835  (2022)

\bibitem{desai2023noise2recon}
Desai, A.D., Ozturkler, B.M., Sandino, C.M., Boutin, R., Willis, M., Vasanawala, S., Hargreaves, B.A., R{\'e}, C., Pauly, J.M., Chaudhari, A.S.: Noise2recon: Enabling snr-robust mri reconstruction with semi-supervised and self-supervised learning. Magnetic Resonance in Medicine  \textbf{90}(5),  2052--2070 (2023)

\bibitem{feng2022spatiotemporal}
Feng, J., Feng, R., Wu, Q., Zhang, Z., Zhang, Y., Wei, H.: Spatiotemporal implicit neural representation for unsupervised dynamic mri reconstruction. arXiv preprint arXiv:2301.00127  (2022)

\bibitem{feng2014golden}
Feng, L., Grimm, R., Block, K.T., Chandarana, H., Kim, S., Xu, J., Axel, L., Sodickson, D.K., Otazo, R.: Golden-angle radial sparse parallel mri: combination of compressed sensing, parallel imaging, and golden-angle radial sampling for fast and flexible dynamic volumetric mri. Magnetic resonance in medicine  \textbf{72}(3),  707--717 (2014)

\bibitem{feng2020grasp}
Feng, L., Wen, Q., Huang, C., Tong, A., Liu, F., Chandarana, H.: Grasp-pro: improving grasp dce-mri through self-calibrating subspace-modeling and contrast phase automation. Magnetic resonance in medicine  \textbf{83}(1),  94--108 (2020)

\bibitem{haft2024neural}
Haft, P.T., Huang, W., Cruz, G., Rueckert, D., Zimmer, V.A., Hammernik, K.: Neural implicit k-space with trainable periodic activation functions for cardiac mr imaging. In: BVM Workshop. pp. 82--87. Springer (2024)

\bibitem{hammernik2018learning}
Hammernik, K., Klatzer, T., Kobler, E., Recht, M.P., Sodickson, D.K., Pock, T., Knoll, F.: {Learning a variational network for reconstruction of accelerated MRI data}. Magnetic resonance in medicine  \textbf{79}(6),  3055--3071 (2018)

\bibitem{hu2021self}
Hu, C., Li, C., Wang, H., Liu, Q., Zheng, H., Wang, S.: Self-supervised learning for mri reconstruction with a parallel network training framework. In: Medical Image Computing and Computer Assisted Intervention--MICCAI 2021: 24th International Conference, Strasbourg, France, September 27--October 1, 2021, Proceedings, Part VI 24. pp. 382--391. Springer (2021)

\bibitem{huang2021lsnet}
Huang, W., Ke, Z., Cui, Z.X., Cheng, J., Qiu, Z., Jia, S., Ying, L., Zhu, Y., Liang, D.: {Deep low-rank plus sparse network for dynamic MR imaging}. Medical Image Analysis  \textbf{73},  102190 (2021)

\bibitem{huang2023neural}
Huang, W., Li, H.B., Pan, J., Cruz, G., Rueckert, D., Hammernik, K.: Neural implicit k-space for binning-free non-cartesian cardiac mr imaging. In: International Conference on Information Processing in Medical Imaging. pp. 548--560. Springer (2023)

\bibitem{jung2010radial}
Jung, H., Park, J., Yoo, J., Ye, J.C.: Radial k-t focuss for high-resolution cardiac cine mri. Magnetic Resonance in Medicine: An Official Journal of the International Society for Magnetic Resonance in Medicine  \textbf{63}(1),  68--78 (2010)

\bibitem{ke2021learned}
Ke, Z., Huang, W., Cui, Z.X., Cheng, J., Jia, S., Wang, H., Liu, X., Zheng, H., Ying, L., Zhu, Y., et~al.: Learned low-rank priors in dynamic mr imaging. IEEE Transactions on Medical Imaging  \textbf{40}(12),  3698--3710 (2021)

\bibitem{korkmaz2023self}
Korkmaz, Y., Cukur, T., Patel, V.M.: Self-supervised mri reconstruction with unrolled diffusion models. In: International Conference on Medical Image Computing and Computer-Assisted Intervention. pp. 491--501. Springer (2023)

\bibitem{kunz2024implicit}
Kunz, J.F., Ruschke, S., Heckel, R.: Implicit neural networks with fourier-feature inputs for free-breathing cardiac mri reconstruction. IEEE Transactions on Computational Imaging  (2024)

\bibitem{kustner2020isotropic}
K{\"u}stner, T., Bustin, A., Jaubert, O., Hajhosseiny, R., Masci, P.G., Neji, R., Botnar, R., Prieto, C.: Isotropic 3d cartesian single breath-hold cine mri with multi-bin patch-based low-rank reconstruction. Magnetic resonance in medicine  \textbf{84}(4),  2018--2033 (2020)

\bibitem{lingala2011accelerated}
Lingala, S.G., Hu, Y., DiBella, E., Jacob, M.: Accelerated dynamic mri exploiting sparsity and low-rank structure: kt slr. IEEE transactions on medical imaging  \textbf{30}(5),  1042--1054 (2011)

\bibitem{mancu2023self}
Mancu, A., Huang, W., da~Cruz, G.L., Rueckert, D., Hammernik, K.: Self-supervised low-rank plus sparse network for radial mri reconstruction. In: NeurIPS 2023 Workshop on Deep Learning and Inverse Problems (2023)

\bibitem{mildenhall2021nerf}
Mildenhall, B., Srinivasan, P.P., Tancik, M., Barron, J.T., Ramamoorthi, R., Ng, R.: {NeRF: Representing scenes as neural radiance fields for view synthesis}. Communications of the ACM  \textbf{65}(1),  99--106 (2021)

\bibitem{muller2022instant}
M{\"u}ller, T., Evans, A., Schied, C., Keller, A.: Instant neural graphics primitives with a multiresolution hash encoding. ACM transactions on graphics (TOG)  \textbf{41}(4),  1--15 (2022)

\bibitem{Nishimura2010}
Nishimura, D.G.: {P}rinciples of {M}agnetic {R}esonance {I}maging. Lulu.com (2010)

\bibitem{otazo2015low}
Otazo, R., Candes, E., Sodickson, D.K.: {Low-rank plus sparse matrix decomposition for accelerated dynamic MRI with separation of background and dynamic components}. Magnetic resonance in medicine  \textbf{73}(3),  1125--1136 (2015)

\bibitem{pan2024unrolled}
Pan, J., Hamdi, M., Huang, W., Hammernik, K., Kuestner, T., Rueckert, D.: Unrolled and rapid motion-compensated reconstruction for cardiac cine mri. Medical Image Analysis  \textbf{91},  103017 (2024)

\bibitem{pan2024reconstruction}
Pan, J., Huang, W., Rueckert, D., K{\"u}stner, T., Hammernik, K.: Reconstruction-driven motion estimation for motion-compensated mr cine imaging. IEEE Transactions on Medical Imaging  (2024)

\bibitem{pipe1999sampling}
Pipe, J.G., Menon, P.: {Sampling density compensation in MRI: rationale and an iterative numerical solution}. Magnetic Resonance in Medicine: An Official Journal of the International Society for Magnetic Resonance in Medicine  \textbf{41}(1),  179--186 (1999)

\bibitem{qin2018convolutional}
Qin, C., Schlemper, J., Caballero, J., Price, A.N., Hajnal, J.V., Rueckert, D.: {Convolutional recurrent neural networks for dynamic MR image reconstruction}. IEEE transactions on medical imaging  \textbf{38}(1),  280--290 (2018)

\bibitem{qiu2024self}
Qiu, Z., Hu, S., Zhao, W., Sakaie, K., Sun, J.E., Griswold, M.A., Jones, D.K., Ma, D.: Self-calibrated subspace reconstruction for multidimensional mr fingerprinting for simultaneous relaxation and diffusion quantification. Magnetic Resonance in Medicine  \textbf{91}(5),  1978--1993 (2024)

\bibitem{rajiah2023cardiac}
Rajiah, P.S., Fran{\c{c}}ois, C.J., Leiner, T.: Cardiac mri: state of the art. Radiology  \textbf{307}(3),  e223008 (2023)

\bibitem{schlemper2017deep}
Schlemper, J., Caballero, J., Hajnal, J.V., Price, A., Rueckert, D.: {A deep cascade of convolutional neural networks for MR image reconstruction}. In: International conference on information processing in medical imaging. pp. 647--658. Springer (2017)

\bibitem{shea2001coronary}
Shea, S.M., Kroeker, R.M., Deshpande, V., Laub, G., Zheng, J., Finn, J.P., Li, D.: Coronary artery imaging: 3d segmented k-space data acquisition with multiple breath-holds and real-time slab following. Journal of Magnetic Resonance Imaging: An Official Journal of the International Society for Magnetic Resonance in Medicine  \textbf{13}(2),  301--307 (2001)

\bibitem{shen2022nerp}
Shen, L., Pauly, J., Xing, L.: Nerp: implicit neural representation learning with prior embedding for sparsely sampled image reconstruction. IEEE Transactions on Neural Networks and Learning Systems  \textbf{35}(1),  770--782 (2022)

\bibitem{sitzmann2020implicit}
Sitzmann, V., Martel, J., Bergman, A., Lindell, D., Wetzstein, G.: {Implicit neural representations with periodic activation functions}. Advances in Neural Information Processing Systems  \textbf{33},  7462--7473 (2020)

\bibitem{spieker2023deep}
Spieker, V., Eichhorn, H., Hammernik, K., Rueckert, D., Preibisch, C., Karampinos, D.C., Schnabel, J.A.: Deep learning for retrospective motion correction in mri: a comprehensive review. IEEE Transactions on Medical Imaging  (2023)

\bibitem{spieker2024self}
Spieker, V., Eichhorn, H., Stelter, J.K., Huang, W., Braren, R.F., R{\"u}ckert, D., Sahli~Costabal, F., Hammernik, K., Prieto, C., Karampinos, D.C., et~al.: Self-supervised k-space regularization for motion-resolved abdominal mri using neural implicit k-space representations. In: International Conference on Medical Image Computing and Computer-Assisted Intervention. pp. 614--624. Springer (2024)

\bibitem{spieker2023iconik}
Spieker, V., Huang, W., Eichhorn, H., Stelter, J., Weiss, K., Zimmer, V.A., Braren, R.F., Karampinos, D.C., Hammernik, K., Schnabel, J.A.: Iconik: Generating respiratory-resolved abdominal mr reconstructions using neural implicit representations in k-space. In: International Conference on Medical Image Computing and Computer-Assisted Intervention. pp. 183--192. Springer (2023)

\bibitem{uecker2016bart}
Uecker, M., Tamir, J.I., Ong, F., Lustig, M.: The bart toolbox for computational magnetic resonance imaging. In: Proc Intl Soc Magn Reson Med. vol.~24, p.~1 (2016)

\bibitem{wang2021fast}
Wang, X., Uecker, M., Feng, L.: Fast real-time cardiac mri: A review of current techniques and future directions. Investigative Magnetic Resonance Imaging  \textbf{25}(4),  252--265 (2021)

\bibitem{wright2014non}
Wright, K.L., Hamilton, J.I., Griswold, M.A., Gulani, V., Seiberlich, N.: {Non-Cartesian parallel imaging reconstruction}. Journal of Magnetic Resonance Imaging  \textbf{40}(5),  1022--1040 (2014)

\bibitem{yaman2020self}
Yaman, B., Hosseini, S.A.H., Moeller, S., Ellermann, J., U{\u{g}}urbil, K., Ak{\c{c}}akaya, M.: Self-supervised learning of physics-guided reconstruction neural networks without fully sampled reference data. Magnetic resonance in medicine  \textbf{84}(6),  3172--3191 (2020)

\end{thebibliography}




\end{document}